\documentclass[prl,preprint,superscriptaddress,longbibliography]{revtex4-1}
\AtBeginDocument{%
	\newwrite\bibnotes
	\def\bibnotesext{Notes.bib}
	\immediate\openout\bibnotes=\jobname\bibnotesext
	\immediate\write\bibnotes{@CONTROL{REVTEX41Control}}
	\immediate\write\bibnotes{@CONTROL{%
			apsrev41Control,author="08",editor="1",pages="1",title="0",year="1"}}
	\if@filesw
	\immediate\write\@auxout{\string\citation{apsrev41Control}}%
	\fi
}%

\usepackage{graphicx}
\usepackage{hyperref}
\usepackage{soul}
\usepackage{amsmath}
\usepackage{amssymb}
\usepackage[usenames,dvipsnames]{color}
\usepackage{ragged2e}

\begin{document}

\title{On-surface Assembly of Au-Dicyanoanthracene Coordination Structures on Au(111)}

\author{Linghao Yan}
\affiliation{Department of Applied Physics, Aalto University School of Science, PO Box 15100, 00076 Aalto, Finland}

\author{Ilona Pohjavirta}
\affiliation{Department of Applied Physics, Aalto University School of Science, PO Box 15100, 00076 Aalto, Finland}

\author{Benjamin Alldritt}
\affiliation{Department of Applied Physics, Aalto University School of Science, PO Box 15100, 00076 Aalto, Finland}

\author{Peter Liljeroth}
\email{Email: peter.liljeroth@aalto.fi}
\affiliation{Department of Applied Physics, Aalto University School of Science, PO Box 15100, 00076 Aalto, Finland}

\date{\today}
\begin{abstract}
On-surface metal-organic coordination provides a promising way for synthesizing different two-dimensional lattice structures that have been predicted to possess exotic electronic properties. Using scanning tunneling microscopy (STM) and spectroscopy (STS), we studied the supramolecular self-assembly of 9,10-dicyanoanthracene (DCA) molecules on the Au(111) surface. Close-packed islands of DCA molecules and Au-DCA metal-organic coordination structures coexist on the Au(111) surface. 
Ordered DCA$_{3}$Au$_{2}$ metal-organic networks have a structure combining a honeycomb lattice of Au atoms with a kagome lattice of DCA molecules. Low-temperature STS experiments demonstrate the presence of a delocalized electronic state containing contributions from both the gold atom states and the lowest unoccupied molecular orbital of the DCA molecules. These findings are important for the future search of topological phases in metal-organic networks combining honeycomb and kagome lattices with strong spin-orbit coupling in heavy metal atoms.
\end{abstract}
\maketitle

It is well-known that the exciting electronic properties of graphene are intimately linked to its honeycomb lattice with a two-atom unit cell \cite{CastroNeto2009TheGraphene}. This results in the formation of Dirac cones in the band structure and the linear dispersion around the K points (at the corners of the Brillouin zone). This is a generic property of any honeycomb lattice and has sparked interest in ``artificial graphene'': engineered systems that have the same structure  \cite{Gomes2012DesignerGraphene,Wang2014ManipulationGas,Paavilainen2016-artificial,Polini2013ArtificialPhotons,Montambaux2018review}. There are other lattice geometries that have the potential to host exotic electronic phases. For example, the kagome lattice has the same Dirac band structure as the honeycomb lattice, but with an additional flat band \cite{Leykam2018ArtificialExperiments} pinned to the top (or bottom) of the Dirac band. Furthermore, these systems can be driven into topological phases by adding spin-orbit coupling. This opens gaps at the band crossings (the Dirac points) which host topological states in finite structures \cite{Kane2005Z2Effect,Hasan2010Colloquium:Insulators,Qi2011TopologicalSuperconductors}.

Metal-organic structures have been synthesized on surfaces following the concepts of supramolecular coordination chemistry \cite{Lin2009Surface-confinedChemistry,Barth2009FreshChemistry}. Their architectures depend on the chemistry of the metal centres with organic linkers and on their interactions with the surface \cite{Barth2007MolecularSurfaces}. Over the past two decades, metal-organic networks with various lattice structures have been fabricated using different combinations of metal atoms and organic molecules. In addition to fabricating \emph{e.g.} simple square geometry, metal-organic networks with honeycomb and kagome lattices have been formed \cite{Dong2016Self-assemblySurfaces}. These hold promise for hosting exotic band structures, especially when combined with heavy metal atoms. In fact, there are several recent predictions based on \emph{ab initio} modelling suggesting that honeycomb and kagome metal-organic networks could host exotic quantum phases, for example, topological insulators \cite{Wang2013OrganicLattices,Wang2013PredictionInsulator,Liu2013FlatFramework, Zhang2016IntrinsicLattices, Zhang2017TheoreticalFramework}. However, most of the metal-organic networks that have been obtained are using 3d transition metals, with only a few reports on heavy metals which can provide strong spin-orbit coupling \cite{Shi2009Porphyrin-basedSurface,Lyu2015On-surfaceStructures,Song2017Self-AssemblySurfaces,Yan2018StabilizingNetworks,Hrvoje2018}. 

Au has an atomic number of 79, making it one of the heaviest of the metal atoms; it possesses a strong spin-orbit coupling and hence, presents an interesting option for metal-organic networks. Au has been found to form two-fold coordination with alkanethiolate \cite{Maksymovych2006Gold-Adatom-MediatedSurface,Voznyy2009TheAu111}, 
phenylthiolate \cite{Maksymovych2008AuAdatomsinSelf-AssemblyofBenzenethiolontheAu111Surface}, 
pyridyl \cite{Shi2009Porphyrin-basedSurface,Song2017Self-AssemblySurfaces}, isocyano \cite{Boscoboinik2010One-dimensionalSurfaces,Zhou2011AdsorptionSurface,Kestell2014DeterminationGold,Feng2014Self-CatalyzedSurfaces,Zhou2015CharacterizationMicroscopy}, and cyano groups \cite{Meyer2015TuningNanostructures,Faraggi2012BondingMolecules}. Three-fold coordination has been reported with the cyano  \cite{Gottardi2014Cyano-FunctionalizedInteractions,Meyer2015TuningNanostructures} and pyridyl groups \cite{Song2017Self-AssemblySurfaces}. 
So far, the Au-coordinated networks are only found with pyridyl group using relatively long linker molecules \cite{Shi2009Porphyrin-basedSurface,Song2017Self-AssemblySurfaces}. Here, we report a scanning tunneling microscopy (STM) and spectroscopy (STS) study of the self-assembly of Au-9,10-dicyanoanthracene (DCA) metal-organic structures on a Au(111) surface. We use the cyano functional group as the building blocks to coordinate with Au. The self-assembled Au-DCA metal-organic structures and close-packed DCA molecules coexist on the Au(111) surface. Au is coordinated in a three-fold Au-DCA motif, which resembles the previously reported Cu-DCA \cite{Pawin2008AExcess,Zhang2014ProbingNetwork} and Co-DCA bonding motifs \cite{Kumar2018Two-DimensionalFrameworks}. Metal-organic networks with DCA$_{3}$Au$_{2}$ stoichiometry are formed through a honeycomb lattice of Au atoms with a kagome lattice of DCA molecules. We found delocalized electronic states in the ordered networks originating from both the Au atoms and the lowest unoccupied molecular orbital (LUMO) of the DCA molecules. These experiments are an important step towards realizing metal-organic networks with heavy metal atoms with the predicted exotic electronic ground states. 


Sample preparation and STM experiments were carried out in an ultrahigh vacuum system with a base pressure of $\sim 10^{-10}$ mbar. The (111)-terminated gold single crystal was cleaned by repeated cycles of Ne$^{+}$ sputtering at 0.75\,kV and annealing to 870~K. 9,10-dicyanoanthracene (DCA, Sigma Aldrich) molecules were thermally evaporated from a resistively heated aluminum oxide crucible at 330~K onto the clean Au(111) substrate held at $\sim200$~K. In addition, we tested the effect of annealing the samples after deposition (see following). Subsequently, the samples were inserted into the low-temperature STM (Createc GmbH) and all subsequent experiments were performed at $T=5$ K. STM images were taken in the constant current mode. d$I$/d$V$ spectra were recorded by standard lock-in detection while sweeping the sample bias in an open feedback loop configuration, with a peak-to-peak bias modulation of 20 mV at a frequency of 526 Hz. All of the STM images were processed by Gwyddion software \cite{Necas2012Gwyddion:Analysis}.

\begin{figure}
    \centering
    \includegraphics[width=\textwidth]{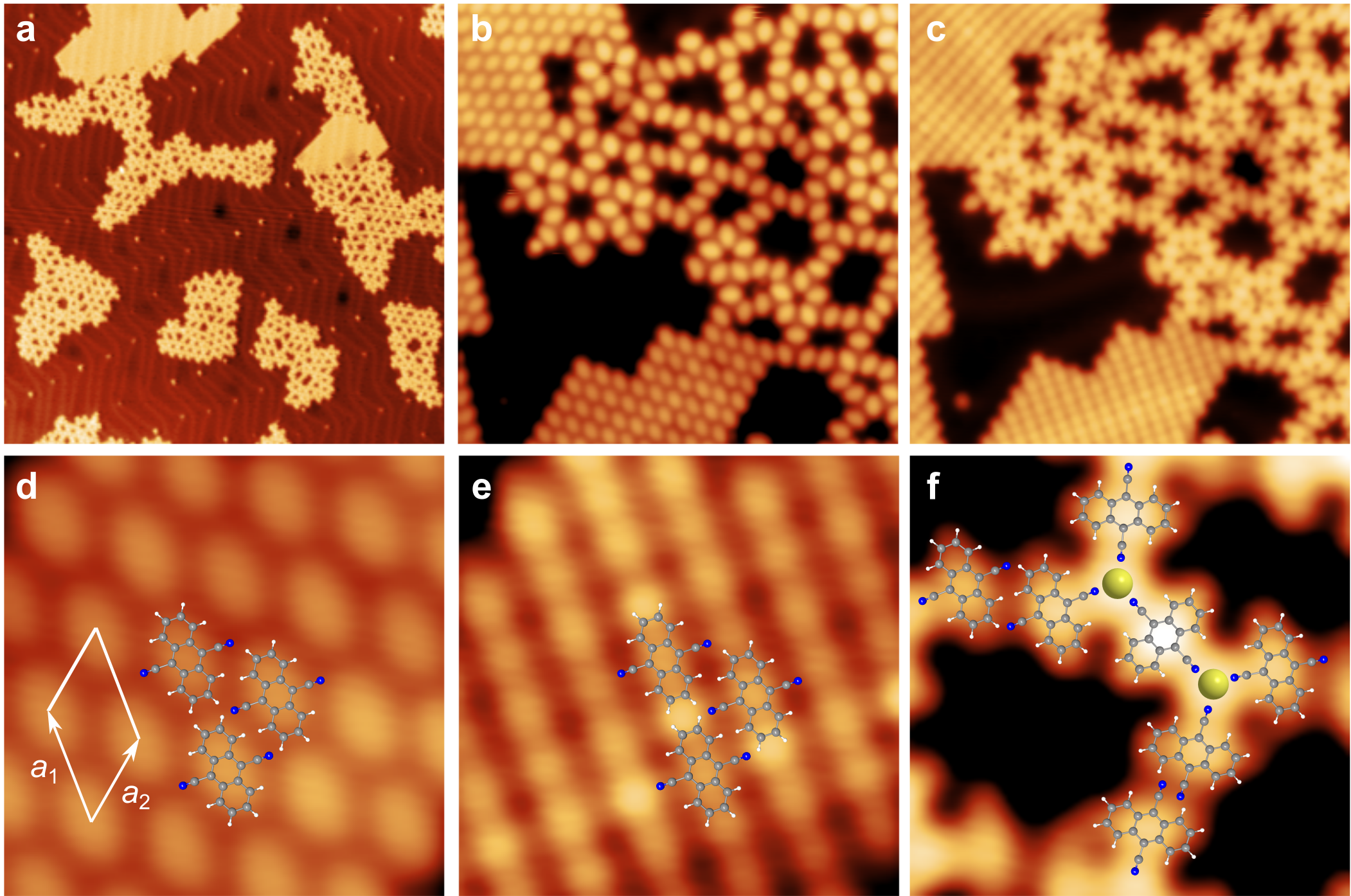}
    \caption{
(a) large-area ($100\times100$ nm$^2$, set-point 1 V/0.1 nA) and (b,c) zoomed-in STM images ($20\times20$ nm$^2$) of DCA supramolecular structures on Au(111) (set-points 0.4 V/0.1 nA (b) and 2.0 V/0.1 nA (c)). (d-f) High resolution STM images ($4\times4$ nm$^2$) with overlaid molecular structures of DCA close-packed phase (set-points 0.4 V/0.1 nA (d) and 2.0 V/0.1 nA (e)) and (f) a mixed phase of Au-DCA coordination motif and hydrogen bonded supramolecular structures (set-point 2.0 V/0.1 nA).
}
    \label{fig:1}
\end{figure}

Fig.~1a shows a large-area STM image of DCA on a Au(111) surface deposited at a temperature of $\sim200$~K. There are two dominant phases with dense, close-packed molecular islands coexisting with a porous structure. These phases can be more clearly seen in Fig.~1b, where it is clear that the dense structure corresponds to the close-packed DCA molecules, while the porous phase corresponds to various self-assembled Au-DCA structures. From Fig.~1c we can see that with a higher sample bias (2 V), there is strong electronic contrast both in the dense and porous structures. Fig.~1d shows a zoom-in on the close-packed phase, where the backbones of the DCA molecules are visible. The unit cell is shown as a white parallelogram in Fig.~1d, where $a_1=1.04 \pm 0.03$ nm, $a_2=0.87 \pm 0.05$ nm and the angle between them is $50^{\circ}\pm0.5^{\circ}$. Similar structure has been found in close-packed DCA molecules on graphene/Ir(111) substrate \cite{Kumar2018Two-DimensionalFrameworks}, while it is in sharp contrast with the square lattice reported on Cu(111) surface \cite{Pawin2008AExcess}, indicating that the DCA molecule-Au(111) substrate interaction is relatively weak. The close-packed DCA molecules are stabilized by multiple C--H...N hydrogen bonds. An image taken at the LUMO resonance of the DCA molecules is shown in Fig.~1e$\colon$ the orbital has a lobe at the each end of the long axis of the anthracene backbone which is consistent with earlier reports \cite{Liljeroth2010Single-MoleculeSTM, Kumar2018Two-DimensionalFrameworks}. The porous structure, as further illustrated with the overlaid molecular and atomic schematics in Fig.~1f, can be understood as a mixed phase of three-fold Au-coordinated DCA together with hydrogen bonded DCA supramolecular structures. The Au--NC bond is estimated to be $2.1 \pm 0.2$ \AA, which is comparable to the value of an earlier report \cite{Meyer2015TuningNanostructures}. 

\begin{figure}
    \centering
    \includegraphics[width=\textwidth]{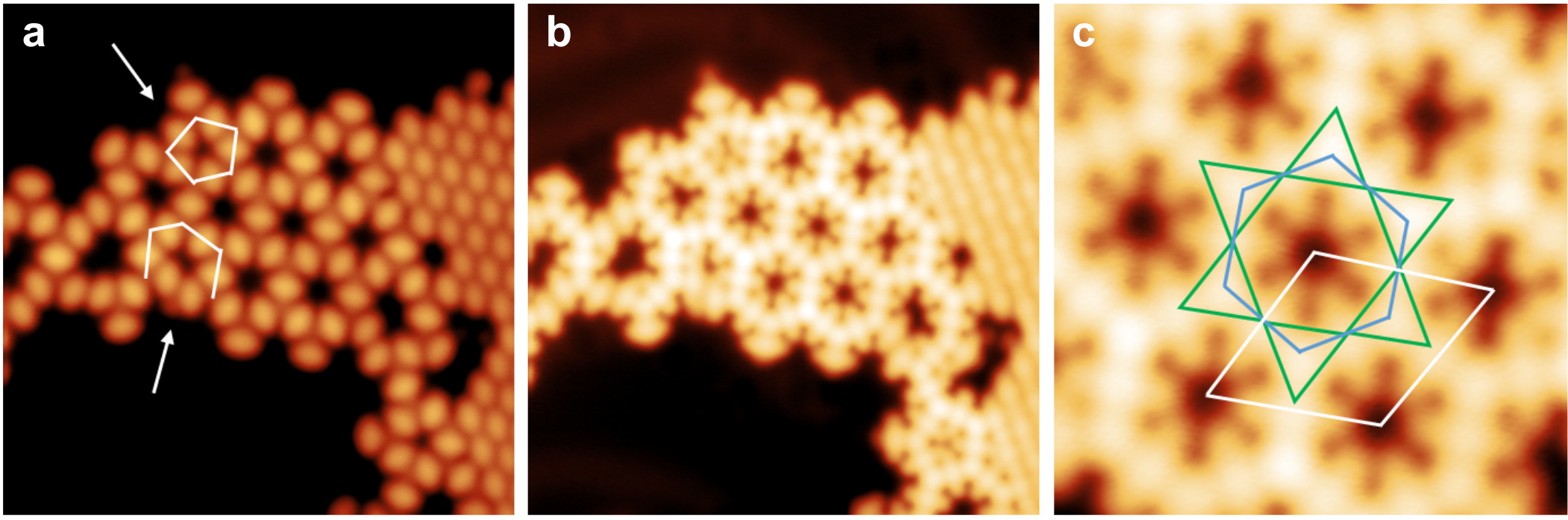}
    \caption{
(a,b) STM images ($15\times15$ nm$^2$) of a small area of highly ordered DCA$_{3}$Au$_{2}$ metal-organic network at two different bias voltages: 1.0 V /10 pA (a) and 1.4 V / 10 pA (b). (c) Zoomed-in STM image with the unit cell of DCA$_{3}$Au$_{2}$ lattice, the honeycomb lattice of Au atoms and the kagome lattice of DCA molecules are presented in white, blue and green lines ($6\times6$ nm$^2$, 1.4 V / 10 pA). Each corner of a green triangle is at a centre of a DCA molecule.
}
    \label{fig:2}
\end{figure}

Despite systematic post-annealing to different temperatures above room temperature, there are no drastic changes in the observed structure until 470 K when there are only individual components left (see Supporting Information). This is consistent with a former work that the cyano-Au coordination bond is formed by the displacement of Au atoms without lifting the herringbone reconstruction \cite{Gottardi2014Cyano-FunctionalizedInteractions}. A relatively highly ordered DCA-Au lattice is found after annealing the sample at 340 K, as demonstrated in Fig.~2. This was the largest ordered island found during the experiments, and we could not detect systematic increase of the size of the highly ordered domains with increasing annealing temperature. We expect that further experiments such as a co-deposition of Au atoms with the DCA molecules or depositing DCA molecules on a heated Au(111) surface may help. Interestingly, at a sample bias of 1.4 V (see Figs.~2b and 2c), the perfect DCA$_{3}$Au$_{2}$ area shows a delocalized electronic contrast all over the backbone of the structure. The defect-free area shown in Fig.~2c exhibits coexisting honeycomb lattice of Au atoms and kagome lattice of DCA molecules, as illustrated by the green and blue lines, respectively. The unit cell is displayed as an overlaid white rhombus in Fig.~2c with a lattice constant of $2.11 \pm 0.05$ nm. As shown in Figs.~2a and 2b, there is an island of close-packed DCA molecules directly on the right side of the defect-free DCA$_{3}$Au$_{2}$ area. The DCA molecules at the boundary are coordinated with Au atoms on one side of the molecule and simultaneously, they seamlessly join the close-packed phase without noticeable phase boundary. The left side of the perfect DCA$_{3}$Au$_{2}$ area terminates with some disordered Au-DCA coordinated structure, where the Au atoms are arranged either in incomplete hexagons or even pentagons as highlighted in Fig.~2a. This suggests that the three-fold Au-DCA coordination motif is very flexible. The flexible Au-DCA coordination bond with the registry of the underlying Au(111) substrate likely hinders the growth of a large-scale, perfect DCA$_{3}$Au$_{2}$ network.

\begin{figure}
    \centering
    \includegraphics[width=.5\textwidth]{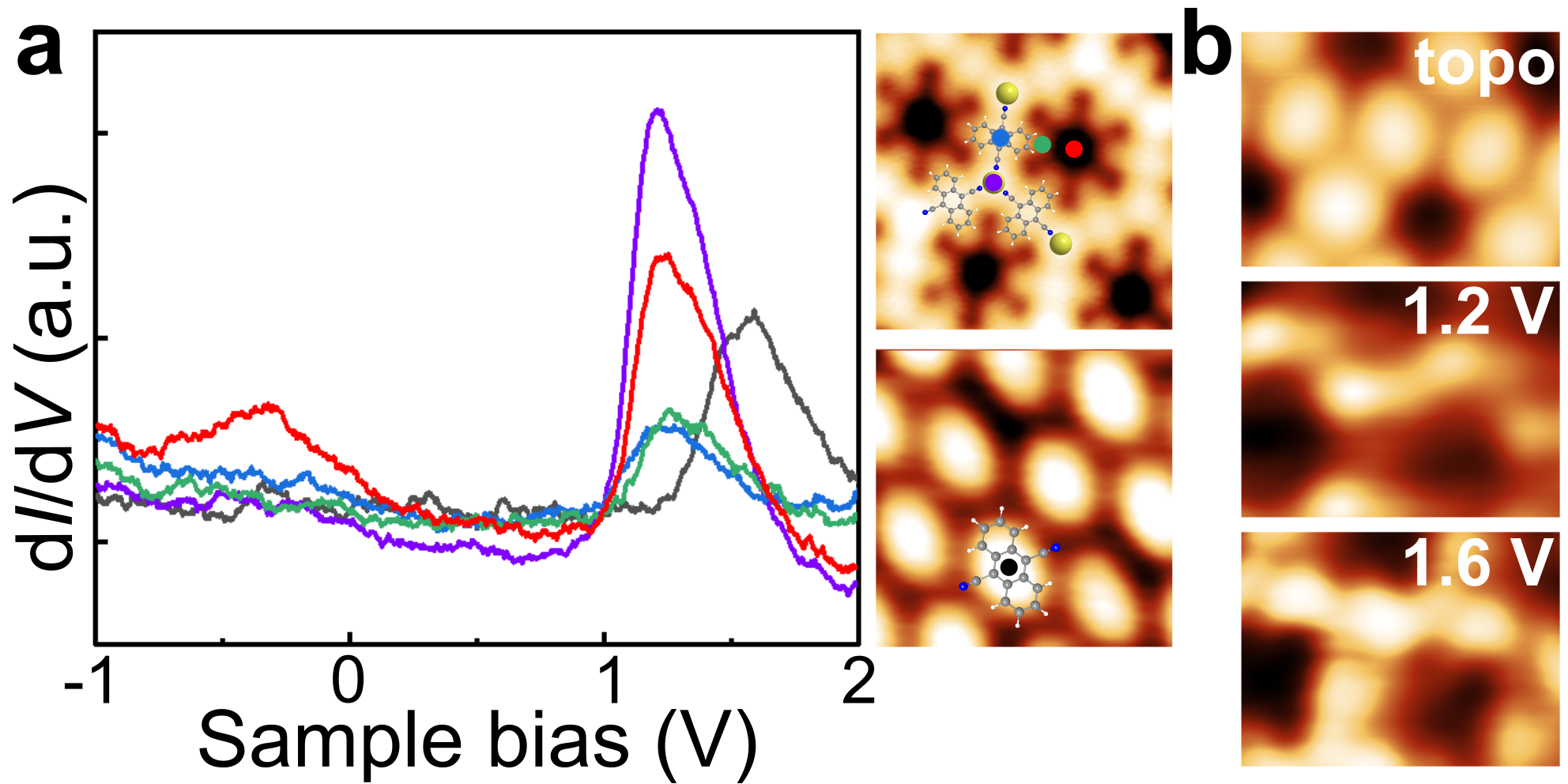}
    \caption{
(a) d$I$/d$V$ spectra recorded on DCA$_{3}$Au$_{2}$ metal-organic network and close-packed DCA island on the positions indicated in the neighbouring STM images. (b) Experimentally recorded topographic image and constant-height d$I$/d$V$ maps ($3.4\times2.5$ nm$^2$) at the bias voltages indicated in the panels. 
}
    \label{fig:3}
\end{figure}

Fig.~3 compares d$I$/d$V$ spectra recorded on close-packed DCA molecules and the DCA$_{3}$Au$_{2}$ network. The d$I$/d$V$ signal is proportional to the local density of states (LDOS) of the sample and allows us to probe the energetic positions of the electronic states in the network as well as their spatial extent \cite{Ervasti2017STM}. As shown in Fig.~3a, d$I$/d$V$ spectrum recorded on a DCA molecule in the close-packed phase shows a broad, asymmetric resonance at 1.6 V corresponding to the LUMO \cite{Liljeroth2010Single-MoleculeSTM,Kumar2018Two-DimensionalFrameworks}. The asymmetric lineshape is caused by phonon replicas: the tunneling electrons can excite molecular vibrations at biases above the elastic peak. The resulting vibronic satellites are separated by the energy of the relevant vibrational mode. The individual satellites are broadened (due to the coupling with the substrate) and cannot be resolved; their sum results in the overall lineshape. \cite{Repp2010CoherentWires,Kumar2018Two-DimensionalFrameworks}.

The d$I$/d$V$ spectra recorded on DCA$_{3}$Au$_{2}$ network show a delocalized peak at 1.2 V that is present all over the intact network. However, if the Au coordination structure is broken, this state is locally suppressed over the missing Au atom (not shown). An additional peak was found in the pore of the network located at -0.32 V which corresponds to the confinement of the surface state of Au(111) substrate \cite{Lobo-Checa2009BandSurface}. The shift of the resonance at positive bias compared to close-packed DCA phase is caused by a change in charge transfer to the DCA due to the Au atoms and changes in the adsorption height of the DCA molecules \cite{Wang2013CooperativeContacts}. In addition to the shift of the energy of the resonance, there are also subtle changes in the lineshape. It seems that electron-vibration coupling strength is reduced and there are multiple components (electronic states) within the resonance. 

Fig.~3b shows the experimentally recorded topographic image and constant-height d$I$/d$V$ maps at biases of 1.2 V and 1.6 V corresponding to the two components visible in the spectra. Comparing the topography and the d$I$/d$V$ maps, we can see that the state on Au atoms is more pronounced at the energy of 1.2 V, while the state at 1.6 V is mainly distributed over the DCA backbones. At 1.2 V, the state originates from both the Au atom states and the down shifted LUMO of DCA molecules. 
This is further evidenced by the d$I$/d$V$ map at 1.6 V. If the other component at 1.6 V was simply due to phonon replicas, similar to the close-packed DCA islands, we would expect no differences in the spatial distribution of the d$I$/d$V$ signal at 1.2 V and 1.6 V \cite{ISI:000180559800042,Repp2010CoherentWires,vanderLit2013SuppressionAtom}. Instead, there is significantly more intensity on the DCA molecules. These findings indicate that the resonance in the d$I$/d$V$ spectrum indeed corresponds to contributions of both Au atom states and the molecular orbitals of DCA.

In conclusion, we studied the self-assembly behaviour of DCA supramolecular structures on Au(111) surface. The samples contained both close-packed DCA phase and a mixed phase Au-DCA coordination motif with hydrogen bonded supramolecular structures. Regions of defect-free DCA$_{3}$Au$_{2}$ coordination network consist of a honeycomb lattice of Au atoms with a kagome lattice of DCA molecules and exhibit delocalized electronic states. It is worth noting that DCA$_{3}$Au$_{2}$ network is predicted to be an intrinsic organic topological insulator \cite{Zhang2016IntrinsicLattices}. We expect that further experimental studies on weakly interacting substrates allowing higher energy resolution spectroscopy will yield more detailed information on the nature of the intrinsic electronic states of these networks \cite{Urgel2015ControllingMonolayer,Ruffieux2016On-surfaceTopology,Kumar2018Two-DimensionalFrameworks}.

\section*{Acknowledgements}
This research made use of the Aalto Nanomicroscopy Center (Aalto NMC) facilities. We acknowledge support from the European Research Council (ERC-2017-AdG no.~788185 ``Artificial Designer Materials'') and Academy of Finland (Academy projects no.~311012 and 314882, and Academy professor funding no.~318995 and 320555).

\section*{Conflict of interest}
The authors declare no conflict of interest.

\section*{Keywords}

Supramolecular self-assembly,
on-surface synthesis,
metal-organic coordination, 
scanning tunneling microscopy and spectroscopy,
ultrahigh vacuum.

\bibliography{references}

\newpage

\section*{Table of contents graphic}

\begin{figure}[h!]
    \includegraphics[width=0.8\textwidth]{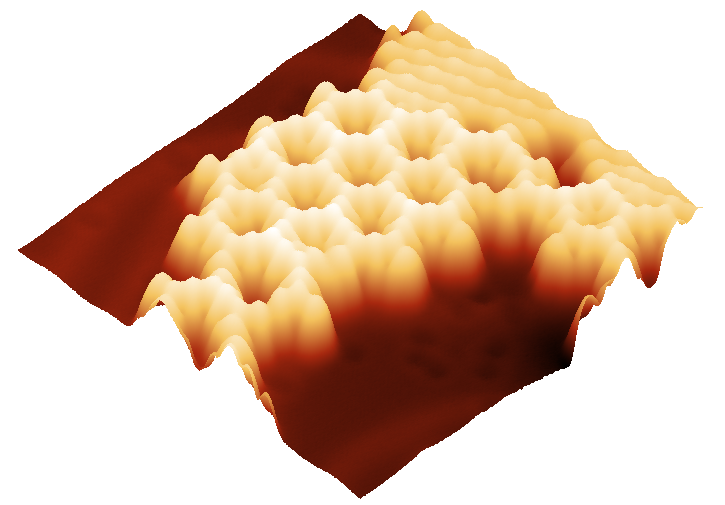}\\
    \justify
    We study the on-surface self-assembly of 9,10-dicyanoanthracene (DCA) molecules on the Au(111) surface to target honeycomb/kagome networks containing heavy metal atoms with strong spin-orbit interaction. We find coexisting close-packed islands of DCA molecules and Au-DCA coordination structures. Ordered DCA$_{3}$Au$_{2}$ metal-organic networks exhibit a delocalized electronic state spread over both the gold atoms and the DCA molecules. 
    \label{TOC}
\end{figure}

\end{document}